\documentclass[english,tightenlines,eqsecnum,amsmath,amssymb,aps,prd,superscriptaddress,nofootinbib,showpacs,floats]{article}
\usepackage{amsmath,amsfonts,amssymb,multicol}
\usepackage{graphicx,caption,subcaption}
\usepackage[usenames]{xcolor}
\definecolor{AHZ}{rgb}{0.0,0.3,0.2}
\usepackage[colorlinks,linkcolor=AHZ,citecolor=red,bookmarks]{hyperref}
\usepackage{rotating}
\usepackage[mathscr]{eucal}
\usepackage{graphicx}
\usepackage[T1]{fontenc}
\usepackage{babel}
\usepackage{authblk}
\usepackage{epsfig}
\usepackage{amsmath,amsfonts,amssymb,multicol}
\usepackage{graphicx,caption,subcaption}
\usepackage[usenames]{xcolor}
\definecolor{AHZ}{rgb}{0.0,0.3,0.2}
\usepackage[colorlinks,linkcolor=AHZ,citecolor=red,bookmarks]{hyperref}
\usepackage{rotating}
\usepackage[mathscr]{eucal}
\usepackage{graphicx}
\usepackage[T1]{fontenc}
\usepackage{babel}
\usepackage{authblk}
\usepackage{epsfig}
\usepackage{color}
\usepackage{amssymb}
\usepackage{fancyhdr}
\newcommand \tlt {\tilde{T}}
\def\nn{\nonumber\\}
\newcommand{\f}[2]{\frac{#1}{#2}}
\def\be{\begin{equation}}
\def\ee{\end{equation}}
\def\bea{\begin{eqnarray}}
\def\eea{\end{eqnarray}}
\def\Nst{N_{\rm stream}}
\def\mst{m_{\rm stream}}
\def\vst{v_{\rm stream}}
\def\sst{\sigma_{\rm stream}}
\def\kpc{~{\rm kpc}}
\def\gyr{~{\rm Gyr}}
\def\kms{~{\rm km/s}}
\def\neff{n_{\rm eff}}
\def\msun{M_{\odot}}
\def\bwt{\begin{widetext}}
\def\ewt{\end{widetext}}
\def\bk{\boldsymbol{k}}
\def\bx{\boldsymbol{x}}
\newcommand{\vi}{\textcolor{cyan}}
\newcommand{\rc}{\textcolor{red}}
\newcommand{\bl}{\textcolor{blue}}
\newcommand{\gr}{\textcolor{green}}

\usepackage{amsmath}

\usepackage{amssymb}

\begin{document}

\title{Stability of the Einstein static Universe in $f(R,T)$ gravity}
\author[1]{Hamid Shabani,\thanks{h.shabani@phys.usb.ac.ir}}
\author[2]{Amir Hadi Ziaie,\thanks{ah.ziaie@gmail.com}}

\affil[1]{Department of Physics, University of Sistan and Baluchestan, Zahedan, Iran}
\affil[2]{Department of Physics, Kahnooj Branch, Islamic Azad University, Kerman, Iran}
\renewcommand\Authands{ and }
\maketitle
\begin{abstract}
The Einstein static (ES) universe has played a major role in various emergent scenarios recently proposed in order to cure the problem of initial singularity of the standard model of cosmology. In the herein model, we study the existence and stability of ES universe in the context of $f(R,T)$ modified theories of gravity. Considering specific forms of $f(R,T)$ function, we seek for the existence of solutions representing  ES state. Using dynamical system techniques along with numerical analysis, we find two classes of solutions: the first one is always unstable of the saddle type while the second is always stable so that its dynamical behavior corresponds to a center equilibrium point. The importance of the second class of solutions is due to the significant duty they have in constructing non-singular emergent models in which the universe could have experienced past-eternally, a series of infinite oscillations about such an initial static state after which, it enters through a suitable physical mechanism, to an inflationary era. Considering specific forms for the functionality of $f(R,T)$, we show that this theory is capable of providing cosmological solutions which admit emergent universe (EU) scenarios. We also investigate homogeneous scalar perturbations for the mentioned models. The stability regions of the solutions are parametrized by a linear equation of state (EoS) parameter and other free parameters that will be introduced for the models. Our results suggest that modifications in $f(R,T)$ gravity would lead to stable solutions which are unstable in $f(R)$ gravity model.
\end{abstract}
\section{Introduction}\label{sec:intro}
In 1917, Einstein put forward an important exact spacetime which is recognized as the first relativistic cosmological model, i.e., the ES universe; a static Friedmann-Robertson-Walker (FRW) model with positive spatial curvature sourced by a perfect fluid and a cosmological constant \cite{ESU}. It was static because it seemed natural to assume that the real universe is static at that time, i.e., qualitatively unchanging in its large scale structure\footnote{For a historical review we consult the reader to see \cite{hreviewESU}.}. However, though this initially appeared to be a reasonable model of a static universe, it was latter shown that ES universe was unstable under small homogeneous and isotropic perturbations around equilibrium state \cite{saeddi}. Since then, ES model has been widely realized to be unstable with respect to gravitational collapse or expansion. It has been also shown that ES universe is always neutrally stable under small inhomogeneous vector and tensor perturbations and also against adiabatic scalar density inhomogeneities as long as the sound speed satisfies\footnote{Subsequent investigations have shown that the ES universe maximizes the entropy for an equation of state with the mentioned value for the sound speed \cite{ESEntropy}} $c_s^2>1/5$, and unstable otherwise \cite{baelltsm}. Moreover, stability of ES universe against the Bianchi type-IX spatially homogeneous perturbations has been studied for various types of matter fields \cite{babianix} and it is found that the ES universe is unstable against such perturbations.
\par One of the most fundamental and ancient questions in standard cosmology is whether the universe has begun from a definite origin or whether it is past-eternal. In the past decades, this question has led to serious discussions based upon the knowledge of general theory of relativity (GR) and modern cosmology.  Recent accumulation of high resolution cosmological observations is compatible with the so called standard model of cosmology which includes number of interesting features. In addition to predicting that the universe is passing through an accelerated phase of expansion \cite{acceluni}, this model also admits an initial singularity that most of the physicists refer to it as the big-bang singularity. Indeed, under general physical circumstances on the matter content of the universe, GR equations imply that the present expanding phase must be preceded by a singular state of the universe where the physical quantities such as energy density and spacetime curvature diverge and the classical framework of GR breaks down \cite{hawpensin}. In order to remedy this shortcoming, a huge amount of work and effort have been recently devoted to construct cosmological models which are non-singular and/or past-eternal. Work along the former model has revealed that non-singular bouncing cosmologies could appear in various scenarios with matter fields violating positive energy conditions like in the quintom bounce \cite{quibounce}, the ghost condensate bounce \cite{ghconbounce} and the Galileon bounce \cite{Galbounce} models or in the modified gravity theories such as string inspired gravity \cite{stringbounce}, Horava gravity \cite{horgrabounce}, non-relativistic gravity \cite{nonrelgrabounce}, gravity in the presence of torsion \cite{torgrabounce}, nonlocal gravity \cite{nonlocbounce}, brane world scenarios \cite{bbounce} and loop quantum cosmology \cite{lqcbounce} (see also \cite{revbounce} for a recent review). Based on the latter model, the search for singularity free cosmological models within the framework of GR has led to development of the so-called EU scenario \cite{Ellisemergent}. In this scenario, the universe is initially in a past-eternal ES state with spatially positive curvature and then eventually evolves to a subsequent inflationary phase. This cosmological model has several remarkable features: there is no initial singularity or \lq\lq{}beginning of time\rq\rq{}; the universe is ever existing and it tends to a static universe in the past infinity rather than originating from a big bang singularity. The universe is isotropic and homogeneous at large scales and may contain exotic matter \cite{Debnath-exotic}. There is also no horizon problem, nor is there a quantum gravity regime (as the model claims) since the scale of curvature always considerably exceeds the Planck scale, so that the space-time may be treated
as a classical entity.
\par Though the main component for constructing emergent scenarios is the ES solution, the original model does not appear to be successful in solving the singularity problem since there is no stable ES solution in classical GR. In other words, owing to the existence of perturbations, such as quantum fluctuations \cite{qflustab}, it is too difficult for the universe to settle down for a long time in such an initial static state \cite{saeddi,ESEntropy,baelltsm}. However, it is a general belief that in its earliest stages, the universe is presumably under extreme physical conditions so that new effects, such as those coming from quantization of gravity, modifications of GR theory or even other new physics, may become significant. As a matter of fact, dealing with cosmological equations of modified gravity theories may leave us with many new static solutions, whose stability properties would crucially depend on the details of the theory. Therefore, it is expected that the outcomes are substantially different from those of the classical ES solution within the GR framework. Thus, it is reasonable to extend the study of ES universe beyond the Einstein gravity. In this regard, stability of the ES universe has been examined in various cosmological settings among which we quote: static cosmological models constructed in brane world models \cite{branesta,branesta1}, modified gravity theories \cite{modifiedgsta,modifiedgsta1}, scalar-fluid theories \cite{SFstability} and loop quantum gravity \cite{lqgsta}. Work along this line has been carried out by considering different types of matter such as, the effects of vacuum energy \cite{vacuumestatic}, non-constant pressure \cite{nonconsprestatic} and a non-interacting mixture of isotropic radiation and a ghost scalar field \cite{nonmixradgoststa}. The first study of ES universe and its stability in $f(R)$ theory of gravity can be found in \cite{fRStability,fRStability1}. In the latter work, it is shown that in contrast to classical GR, the modified ES universe can be stabilized against homogeneous perturbations in the context of two well-known forms of $f(R)$ with a positive cosmological constant and matter content described by a barotropic perfect fluid with equation of state (EoS) $p=w\rho$. However, subsequent work on the ES universe within the context of generic $f(R)$ models has shown that Einstein static solutions are always unstable against homogeneous or inhomogeneous perturbations \cite{genericfRstability}.
\par Recently, a kind of modified theory of gravity has been developed as $f(R,T)$ gravity which was first introduced in \cite{frtHarko}. This new proposal for modifying the gravity has been widely studied within various contexts such as thermodynamics~\cite{frttd1,frttd2,frttd3,frttd4}, energy conditions~\cite{frtec1,frtec2,frtec3}, cosmological solutions from dynamical system point of view  ~\cite{phsp1,phsp2}, anisotropic cosmology~\cite{frtani1,frtani2,frtani3}, wormhole solution~\cite{frtworm}, scalar perturbations~\cite{frtsp}, cosmology of non-interacting Chaplygin gas~\cite{frtChap} and some other studies such as, $f(R,T)$ gravity in higher dimensions~\cite{frtoth1}, the effects of matter-curvature coupling on the distribution of matter configuration for a self-gravitating spherical body~\cite{frtoth2} and dark matter effects in spiral galaxies~\cite{frtoth3}. This theory extends $f(R)$ gravity by including the trace of energy-momentum tensor (EMT), in addition to the Ricci curvature scalar. The motivation of including the trace of EMT may come from the effects of some exotic fluid, consequences of some unknown gravitational interactions or even quantum effects (conformal anomaly)~\cite{fRTconf}. Motivated by the above discussion, in the present work we investigate the existence of static solutions and their stability in the framework of $f(R,T)$ theory of gravity. This paper is organized as follows: in section~\ref{Field eqs} we present the field equations of $f(R,T)$ gravity and some related definitions and also give a few discussions on the conservation of EMT. In sec~\ref{ES sol}, the ES solution and its stability is investigated under two class of models: class I, which belongs to the $f(R,T)$ models that respect to the conservation of EMT and will be studied in subsection~\ref{sub1}. The second class introduces the models in which the conservation of EMT is relaxed and will be presented in subsections~\ref{sub2} for a pressure-less matter and~\ref{sub3} for a barotropic perfect fluid. For the latter case, we find that depending on the EoS parameter and other model parameters, an EU scenario could arise from the ES universe. In section~\ref{pert}, we investigate the stability of ES universe against homogeneous scalar perturbations and show that the results are consistent with those we shall find in subsections~\ref{sub1} and \ref{sub2}. Finally, in section~\ref{con}, we summarize our results.
\section{Field equations of $f(R,T)$ gravity}\label{Field eqs}
In this section we present the equations of motion for $f(R,T)$ theories of gravity in the presence of radiation and the cold dark matter as the matter contents. This modified gravity model is governed by the action
\begin{align}\label{action}
S=\int \sqrt{-g} d^{4} x \left[\frac{1}{2\chi^{2}} f\Big{(}R, T^{\textrm{(r, c)}}\Big{)}
+L^{\textrm{(total)}} \right],~~~~L^{\textrm{(total)}}\equiv L^{\textrm{(r)}}+L^{\textrm{(c)}},
\end{align}
where $\chi^{2}= 8\pi G$, being the gravitational coupling constant, $L^{{\rm (total)}}$ being the Lagrangian of the total matter and $R$, $T^{\textrm{(r, c)}}$ and $L^{\textrm{(total)}}$ are the Ricci curvature scalar, the trace of EMT of radiation and cold dark matter ($T^{\textrm{(r, c)}}_{\mu \nu}\equiv T^{\textrm{(r)}}_{\mu \nu}+
T^{\textrm{(c)}}_{\mu \nu}$) and the Lagrangian of whole matter fields,
respectively. We take these two types of matter as the only sources for gravitational interaction. It is worth noticing that, the equations we shall derive in the present section will be employed in subsections \ref{sub1} and \ref{sub2}. However, in subsection \ref{sub3}, we rewrite some of them for a single perfect fluid with linear EoS parameter, $p=w\rho$. The superscripts $({\rm r,c})$ stand for the radiation and pressure-less matter fields and $g$ is the determinant of
the metric. We work in the units in which $c=1$. The energy-momentum tensor
$T_{\mu \nu}^{\textrm{(r, c)}}$ is defined as
\begin{align}\label{Euler-Lagrange}
T_{\mu \nu}^{\textrm{(r, c)}}\equiv-\frac{2}{\sqrt{-g}}
\frac{\delta\left[\sqrt{-g}(L^{\textrm{(r)}}+L^{\textrm{(c)}})
\right]}{\delta g^{\mu \nu}}.
\end{align}
where $L^{\textrm{(r)}}$ and $L^{\textrm{(c)}}$ are the Lagrangians of the
radiation and the cold dark matter. The field equations for $f(R,T)$
gravity can be derived via varying action (\ref{action}) with respect to the metric field and are given as~\cite{frtHarko}
\begin{align}\label{fRT field equations}
F(R,T) R_{\mu \nu}-\frac{1}{2} f(R,T) g_{\mu \nu}+\Big{(} g_{\mu \nu}
\square -\triangledown_{\mu} \triangledown_{\nu}\Big{)}F(R,T)\\
=\Big{(}\chi^{2}-{\mathcal F}(R,T)\Big{)}T_{\mu \nu}-\mathcal {F}(R,T)\mathbf
{\Theta_{\mu \nu}},\nonumber
\end{align}
where
\begin{align}\label{theta}
\mathbf{\Theta_{\mu \nu}}\equiv g^{\alpha \beta}\frac{\delta
T_{\alpha \beta}}{\delta g^{\mu \nu}},
\end{align}
and for simplicity we have defined the following functions for derivatives of $T$ and $R$, as
\begin{align}\label{f definitions1}
{\mathcal F}(R,T) \equiv \frac{\partial f(R,T)}{\partial T}~~~~~
~~~~~\mbox{and}~~~~~~~~~~
F(R,T) \equiv \frac{\partial f(R,T)}{\partial R}.
\end{align}
Note that, since $T^{{\rm (r)}}=0$, only $T^{{\rm (c)}}$ can appear in the function $f(R,T)$.
Therefore, the superscript ({\rm c}) will be dropped hereafter unless it is needed. We assume that the universe is filled with a perfect fluid that evolves in a spatially non-flat Friedmann--Lema\^{\i}tre--Robertson--Walker (FLRW) spacetime whose line element can be parametrized as
\begin{align}\label{metricFRW}
ds^{2}=-dt^{2}+a^{2}(t) \Big{(}\frac{dr^{2}}{1-kr^2}+r^{2}d\Omega^2\Big{)}.
\end{align}
Substituting the above metric into the field equations (\ref{fRT field equations}) and taking the radiation and dark matter as the matter sources, we get
\begin{align}\label{first}
&3H^{2}F(R,T)+\frac{1}{2} \Big{(}f(R,T)-F(R,T)R\Big{)}+3\dot{F}
(R,T)H+3\frac{kF(R,T)}{a^2}\nonumber\\
&=\Big{(}\chi^{2} +{\mathcal F}
(R,T)\Big{)}\rho^{\textrm{(c)}}+\chi^{2}\rho^{\textrm{(r)}},
\end{align}
and
\begin{align}\label{second}
&2F(R,T) \dot{H}+\ddot{F} (R,T)-\dot{F} (R,T)-2\frac{kF(R,T)}{a^2} H\nonumber\\
&=-\Big{(}\chi^{2}+{\mathcal F} (R,T)\Big{)}\rho^{\textrm{(c)}}-\frac{4}{3}\chi^{2}\rho^{\textrm{(r)}},
\end{align}
where $\dot{}\equiv d/dt$. To avoid mathematical complexities and other difficulties we devote the rest
of our work to a simple form for the functionality of $f(R,T)$ as
\begin{align}\label{choice}
f(R,T)=R+h(T),
\end{align}
whence, the field equations (\ref{first}) and (\ref{second}) for a closed universe ($k=1$) can be rewritten as
\begin{align}\label{form-1}
3H^{2}=-\Big{(}\chi^{2}+h'(T)\Big{)}T +\chi^{2} \rho^{(r)}-\frac{h(T)}{2}-\frac{3}{a^{2}},
\end{align}
and
\begin{align}\label{form-2}
2\dot{H}=\Big{(}\chi^{2}+h'(T)\Big{)}T-\frac{4}{3}\chi^{2}\rho^{(r)}+\frac{2}{a^{2}}.
\end{align}

For later applications, we investigate two situations. We first consider the case(s) in which
the conservation of EMT is respected and then proceed with studying those cases for which the conservation of EMT does not
hold. In the first one, the following constraint for multi perfect fluids
can be calculated by applying the Bianchi identity to the field equation (\ref{fRT field equations}) as
\begin{align}\label{constraint1}
\sum_{i=1}^{N}\dot{\mathcal {F}_{i}}(R,T)(\rho_{i}+p_{i})-
\frac{1}{2}\mathcal {F}_{i}(R,T)(\dot{p_{i}}-\dot{\rho_{i}})=0,
\end{align}
where $N$ is the number of perfect fluids. The satisfaction of this constraint
guaranties the conservation of EMT. The above constraint equation
for a pressure-less matter gives\footnote{See \cite{phsp1,phsp2} for more details.}
\begin{align}\label{constraint2}
\dot{\mathcal {F}}(R,T)=\frac{3}{2}H\mathcal {F}(R,T),
\end{align}
where $H$ is the Hubble parameter. Substituting expression (\ref{choice}) into the constraint (\ref{constraint2}) and after a straightforward algebra, we get a specific form for $h(T)$ as
\begin{align}\label{specific}
h(T)=C_{1}\sqrt{|T|}+C_{2},
\end{align}
where $C_{1}$ and $C_{2}$ are constants of integration.

From another side, if we set the conservation of EMT to be relaxed, the Bianchi identity would
lead to the following covariant equation between the function $\mathcal {F}(R,T)$, the EMT and its trace as
\begin{align}\label{relation}
(\chi^{2} +\mathcal {F})\bigtriangledown^{\mu}T^{{\rm (total)}}_{\mu \nu}+
\frac{1}{2}\mathcal {F}\bigtriangledown_{\mu}T+T^{{\rm (total)}}_{\mu \nu}\bigtriangledown^{\mu}\mathcal {F}
-\nabla_{\nu}(p\mathcal{F})=0,
\end{align}
where we have dropped the argument of $\mathcal {F}(R,T)$. Notice that in the
last fourth terms of equation (\ref{relation}), only $T^{{\rm (c)}}$ would appear since the function $\mathcal {F}$ and its derivative are non-zero only for the cold dark matter.
However, In the first term we still have the term $\chi^{2}
\bigtriangledown^{\mu}T^{{\rm (r)}}_{\mu \nu}$ for the radiation part. Therefore, equation (\ref{relation})
can be considered as the sum of two terms; those that are related to the radiation and other terms which are related to the cold dark matter (for $p^{\rm (c)}=0$), which totally must be set to zero. Nevertheless, a simple choice is
\begin{align}
&(\chi^{2} -\mathcal {F})\bigtriangledown^{\mu}T^{{\rm (c)}}_{\mu \nu}+
\frac{1}{2}\mathcal {F}\bigtriangledown_{\mu}T+T^{{\rm (c)}}_{\mu \nu}\bigtriangledown^{\mu}\mathcal {F}
+2\mathcal {F}\bigtriangledown^{\mu}T^{{\rm (c)}}_{\mu \nu}=0,\label{relation-1}\\
&\bigtriangledown^{\mu}T^{{\rm (r)}}_{\mu \nu}=0,\label{relation-2}
\end{align}
which means that, the radiation and the cold dark matter would evolve independently such that
the radiation does follow the conservation of EMT. It can then be easily seen that the radiation density would depend on the scale factor as $\rho^{{\rm (r)}}\varpropto a^{-4}$. However, the evolution of the cold dark matter density would follow the solution of equation (\ref{relation-1}) rather than the usual case in the standard cosmology i.e., $\rho^{{\rm (c)}}\varpropto a^{-3}$.
For the line element (\ref{metricFRW}), equation (\ref{relation-1}) takes the following form
\begin{align}\label{relation-3}
\Big{(}\chi^{2} + \frac{3}{2}\mathcal {F}- \mathcal {F}'\rho^{{\rm (c)}}\Big{)}\dot{\rho}^{{\rm (c)}}+
3H\rho^{{\rm (c)}}\Big{(}\chi^{2} + \mathcal {F}\Big{)}=0.
\end{align}
Thus, having determined the functionality of $h(T)$ in the non-conserved case, it is the above equation that governs the behavior of the cold dark matter density in terms of the scale factor. For example, if we set
\begin{align}\label{case-non}
f(R,T)=R+n\chi^{2} T,
\end{align}
we obtain
\begin{align}\label{ro-non}
\rho^{{\rm (c)}}=\rho_{0}^{{\rm (c)}}a^{-3\gamma},~~~~~\gamma=\frac{2+2n}{2+3n}.
\end{align}

\section{The Einstein static solution, existence and stability}\label{ES sol}
Our attempt here is to bring forward three classes of solutions that can be served as the ES universe models. We deal with the solution (\ref{specific}) in subsection (\ref{sub1}) as the only conserved case. We then proceed to investigate the non-conserved case (\ref{case-non}) in subsection (\ref{sub2}). In these two subsections, radiation and cold dark matter are taken as the whole matter content of universe. Finally, subsection (\ref{sub3}) is devoted to ES solution for a perfect fluid with linear EoS parameter $w=p/\rho$ and thus some required equations will be rewritten from section~\ref{Field eqs}. To illustrate the obtained results we shall present some diagrams.

\subsection{Class I: Conserved  EMT, case i: $f(R,T)=R+C_{1}\sqrt{|T|}$, $w=0$}\label{sub1}
\begin{figure}
\centering
\includegraphics[scale=0.29]{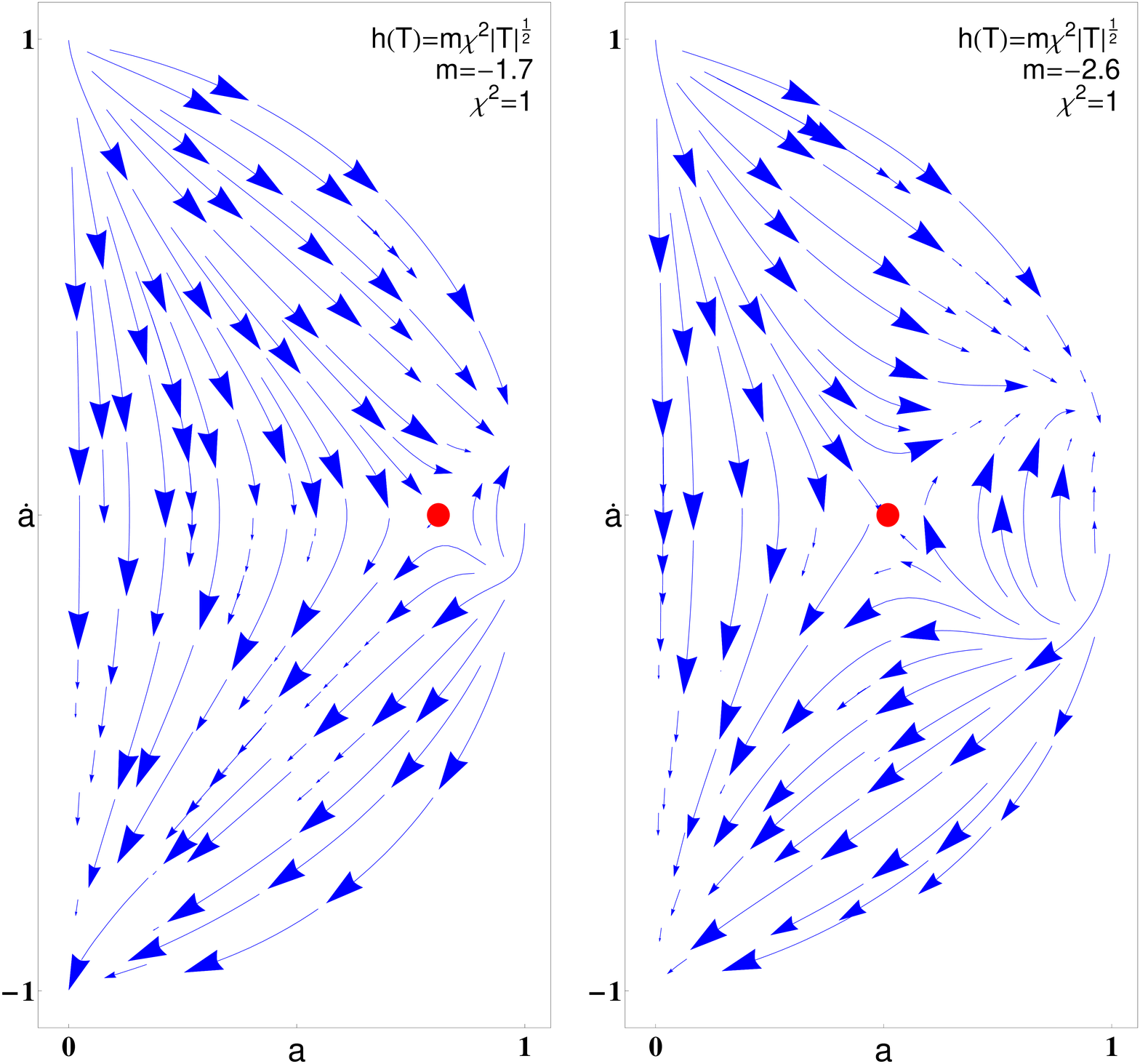}
\caption{The phase space portrait for the conserved case $h(T)=C_{1}\sqrt{|T|}$ in
$(a,\dot{a})$ plane. The red solid circle represents the unstable static solution given at the point $(a^{{\rm (con)}}_{{\rm ES}},0)$.}\label{fig1}
\end{figure}

In this case, equations (\ref{form-1}), (\ref{form-2}) and (\ref{constraint2}) lead to
the following equations
\begin{align}\label{con-1}
3H^{2}=\chi^{2}\rho^{{\rm (c)}}+\chi^{2}\rho^{{\rm (r)}}-h(T)-\frac{3}{a^{2}},
\end{align}
and
\begin{align}\label{con-2}
2\dot{H}=-\chi^{2}\rho^{{\rm (c)}}-\frac{4}{3}\chi^{2}\pi G \rho^{{\rm (r)}}+\frac{1}{2}h(T)+\frac{2}{a^{2}}.
\end{align}
Combining equations (\ref{con-1}) and (\ref{con-2}) leaves us with the following equation for the acceleration of the universe as
\begin{align}\label{acceleration}
\ddot{a}=-\frac{\dot{a}^2+1}{2a}-a\Big{(}\frac{\chi^{2}}{6}\rho^{{\rm (r)}}+\frac{1}{4}h(T)\Big{)},
\end{align}
where for this case we have $T^{{\rm (c)}}=-\rho^{{\rm (c)}}$. The ES solution is given by
the conditions $\dot{a}=0$,~$\ddot{a}=0$ and~$\dot{\rho}=0$. Using these conditions, equation
(\ref{acceleration}) gives
\begin{align}\label{radius}
-\frac{1}{2a}-a\Big{(}\frac{\chi^{2}}{6}\rho^{{\rm (r)}}+\frac{1}{4}h(T)\Big{)}=0.
\end{align}
Now, given the functionality of $\rho^{{\rm (r)}}$ and $\rho^{{\rm (c)}}$ in terms of the scale factor together with determining $h(T)$, equation (\ref{radius}) can be solved for the scale factor of the
ES universe. Substituting for $h(T)=C_{1}\sqrt{|T|}$ (here, we set $C_{2}=0$ as a cosmological constant) and $\rho^{{\rm (r)}}$ into equation (\ref{radius}), we get
\begin{align}\label{radius-1}
-\frac{1}{2a}-\Big{(}\frac{\chi^{2}}{6}\rho^{{\rm (r)}}_{0}a^{-3}+
m\frac{\chi^{2}}{4}\rho^{{\rm (c)}}_{0}a^{-1/2}\Big{)}=0,
\end{align}
where we have set $C_{1}\equiv m\chi^{2}\sqrt{\rho^{{\rm (c)}}_{0}}$. Rewriting equation
(\ref{radius-1}) in terms of the cosmological parameters we have
\begin{align}\label{radius-12}
-\frac{1}{2a}-\frac{H_{0}^{2}}{2}\Big{(}\Omega_{0}^{{\rm (r)}}a^{-3}+
\frac{3m}{2}\Omega_{0}^{{\rm (c)}}a^{-1/2}\Big{)}=0,
\end{align}
where $H_{0}$ is the Hubble constant and $\Omega_{0}^{{\rm (r)}}$ and $\Omega_{0}^{{\rm (c)}}$, are present values for density parameters of radiation and cold dark matter, respectively.
Since $\Omega_{0}^{{\rm (r)}}\ll\Omega_{0}^{{\rm (c)}}$ at the present moment, equation (\ref{radius-12})
can be approximately solved for the radius of curvature of ES universe, as
\begin{align}\label{sol-radius-1}
a_{{\rm ES}}^{{\rm(con)}}\simeq\Big{(}\frac{2}{3m\Omega_{0}^{{\rm(c)}}
H_{0}^{2}}\Big{)}^{2},
\end{align}
where the superscript \lq\lq{}con\rq\rq{} is used for the related parameters of the conserved case
and the solution is valid only for $m<0$. Next, we proceed to examine the stability of
the solution (\ref{sol-radius-1}). To this aim, using the Raychaudhuri equation (\ref{acceleration}), we introduce phase space variables $x=a$ and $y=\dot{a}$ to establish the autonomous system of equations as
\begin{align}\label{auto-1}
&\dot{x}=y,\\
&\dot{y}=-\frac{y^{2}+1}{2x}-\frac{H_{0}^{2}}{2}\Big{(}\Omega_{0}^{{\rm (r)}}x^{-3}+
\frac{3m}{2}\Omega_{0}^{{\rm(c)}}x^{-1/2}\Big{)}.
\end{align}
In terms of these variables, the ES solution corresponds to critical point of the above dynamical system which is given as $x=a_{{\rm ES}}^{\rm (con)}$ and $y=0$. The stability analysis of the solution can be easily performed by finding the eigenvalues of the Jacobian matrix $(J_{ij}=\partial\dot{q}_i/\partial q_j)$ evaluated at the critical point. We then get, after some calculations
\begin{align}\label{eigen-1}
\lambda_{1,2}^{{\rm (con)}}=\pm\frac{1}{8}\Big{(}3m\Omega_{0}^{{\rm(c)}}
H_{0}^{2}\Big{)}^{2}=\pm\frac{1}{2a_{{\rm ES}}^{{\rm(con)}}}.
\end{align}
This means that the fixed point $(x=a_{{\rm ES}}^{{\rm(con)}}, y=0)$ is a saddle point which
is referred to as an unstable one. In such cases, depending on the initial values of the system,
some trajectories in the phase space would approach the fixed point and some others
would get away from it. The phase space portrait for two different values $m=-1.7$
and $m=-2.6$ is drawn in Fig~\ref{fig1}. The red solid circle denotes the equilibrium point $(x=a_{{\rm ES}}^{{\rm(con)}}, y=0)$.
It is seen that, the smaller values for $m$ leads to smaller radius for ES universe.
Note that, in order to show the behavior of trajectories in the phase space more accurately, we have set
$H_{0}^{2}\Omega_{0}^{{\rm(c)}}=1$. Therefore, this class of solutions does not admit a stable ES universe.
~
\subsection{Class II: Non-conserved EMT, case ii: $f(R,T)=R+n\chi^{2}T$, $w=0$}\label{sub2}
\begin{figure}[h]
\includegraphics[scale=0.28]{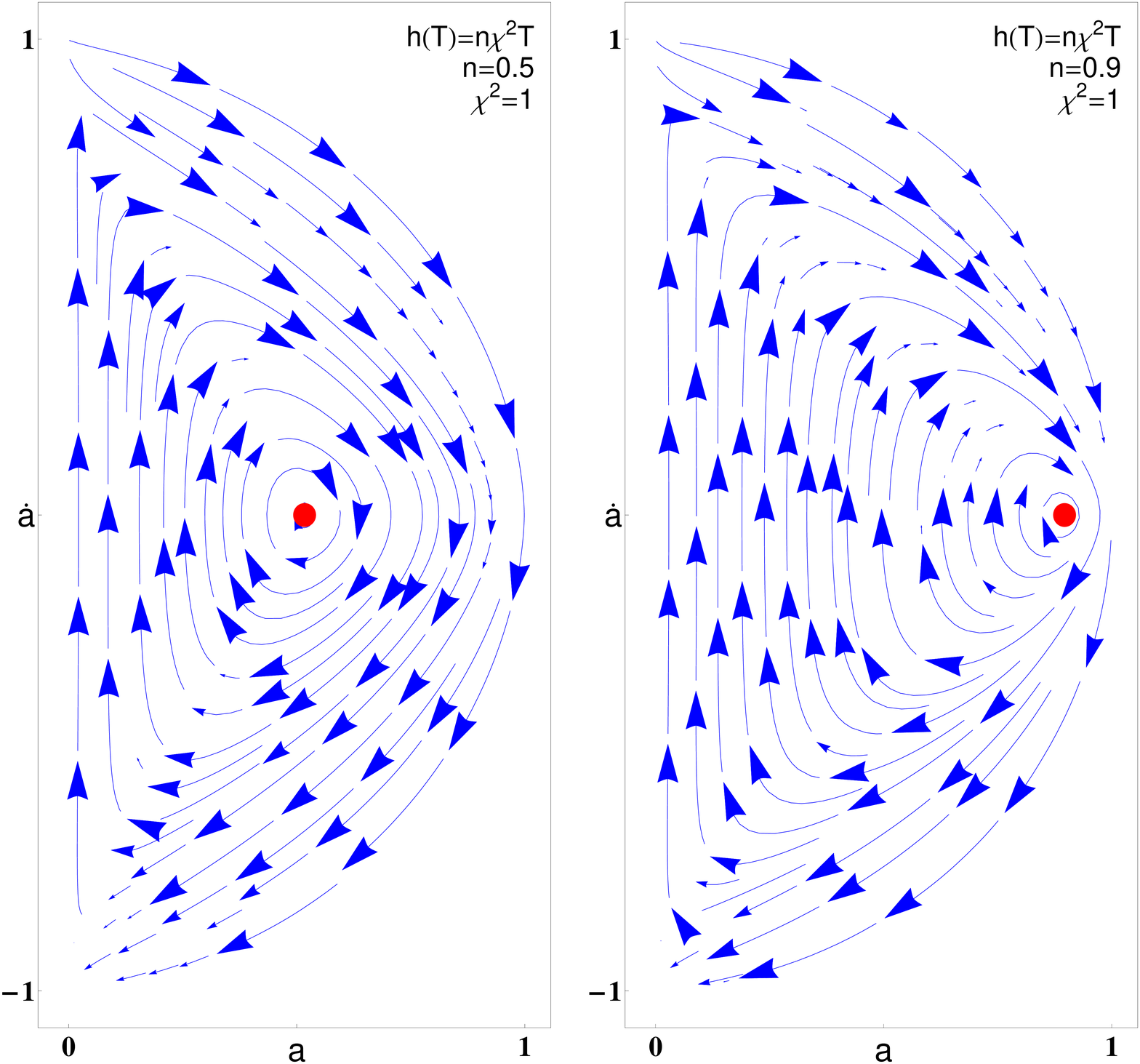}
\caption{\footnotesize {The phase space portrait in $(a,\dot{a})$ plane for the non-conserved case, $h(T)=n\chi^{2}T$. The red solid circle represents the stable static solution.}}
\label{Fig2}
\end{figure}

\begin{figure}[h]
\centering\epsfig{figure=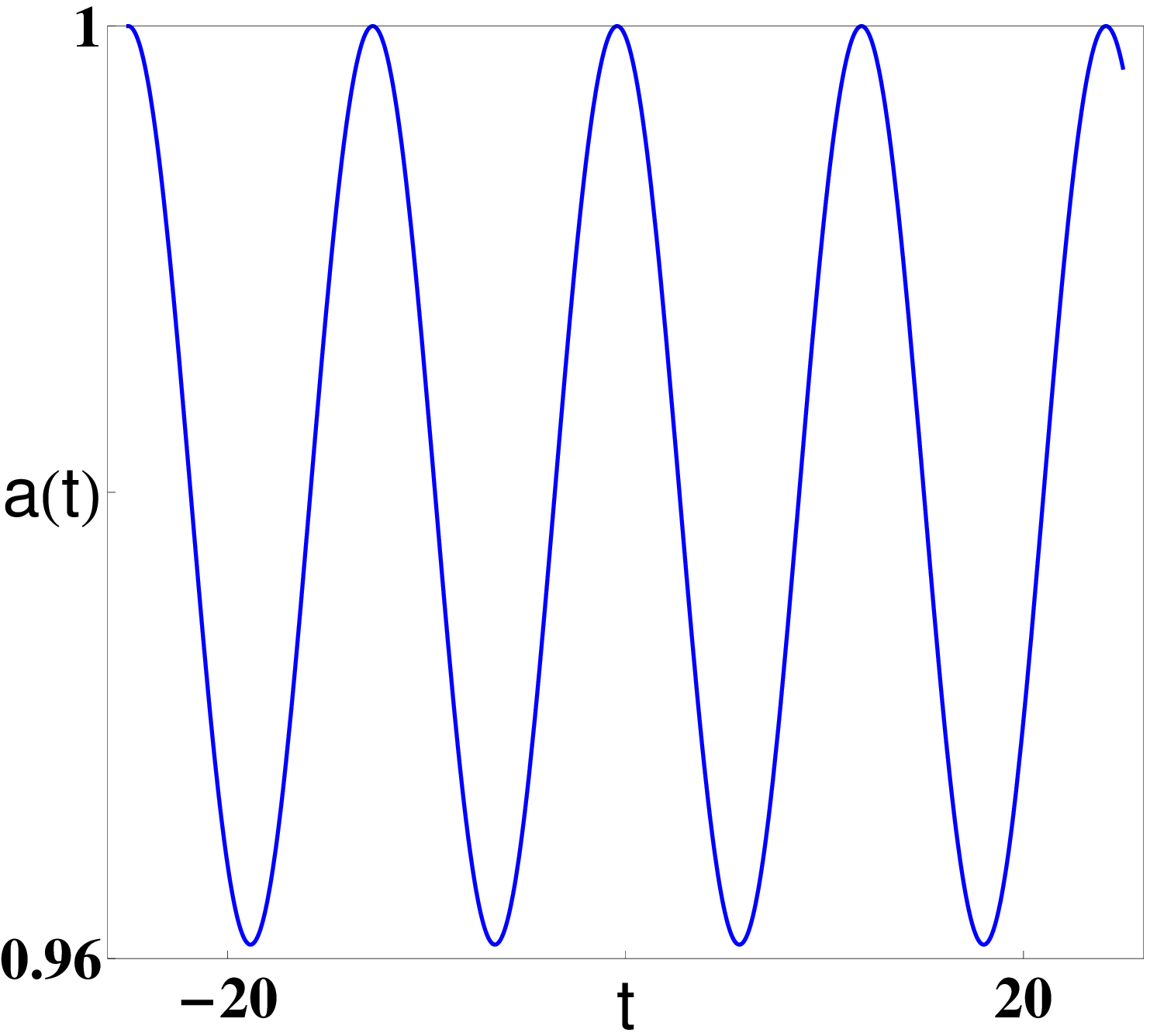,width=5.7cm}\hspace{2mm}
\centering\epsfig{figure=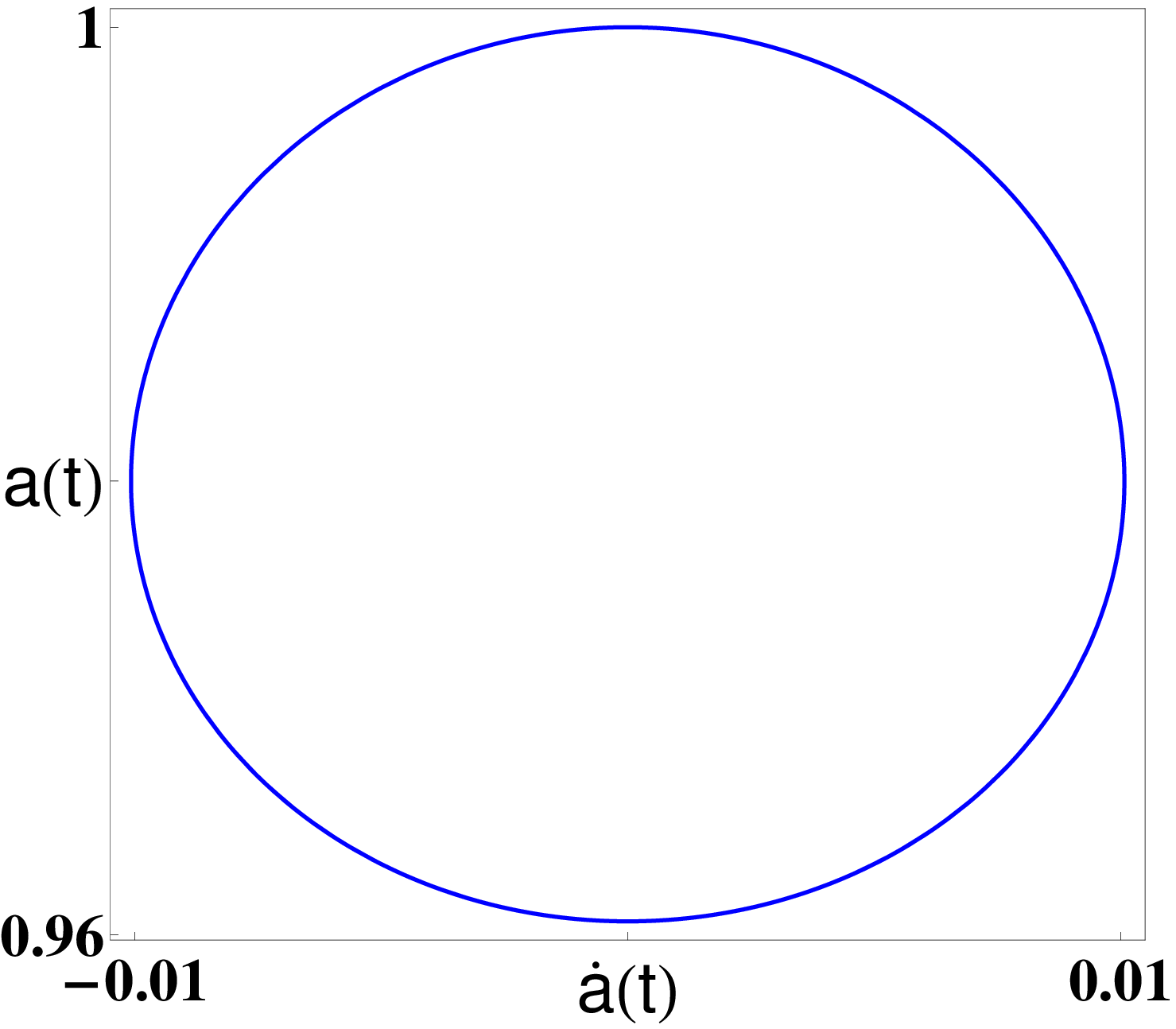,width=5.7cm}
\caption{\footnotesize {Left panel: the behavior of the scale factor with respect to time for Class II solutions. Right panel: the trajectory of the ES universe in the $(\dot{a}, a)$ plane for the same case. The initial values $\dot{a}_i=0$ and $a_i=1$ have been chosen and we have set $n=0.66$.}}
\label{Fig3}
\end{figure}

In this case, using equations (\ref{form-1}) and (\ref{form-2}) for the
non-conserved case and setting $h(T)=n\chi^{2}T$, (where $n$ is a dimensionless parameter) we obtain the following Raychaudhuri equation
in terms of the Hubble constant and matter density parameters as
\begin{align}\label{radius-21}
\ddot{a}=-\frac{1+\dot{a}^{2}}{2a}-\frac{H_{0}^{2}}{2}\Big{(}\Omega_{0}^{{\rm(r)}}a^{-3}-
\frac{3n}{2}\Omega_{0}^{{\rm(c)}}a^{-3\gamma+1}\Big{)}.
\end{align}
Using the fact that $\Omega_{0}^{{\rm(r)}}\ll\Omega_{0}^{{\rm(c)}}$ together with setting $\ddot{a}=\dot{a}=0$, the above equation can be easily solved for the scale factor. The solution is given as
\begin{align}\label{sol-radius-2}
a_{{\rm ES}}^{{\rm(n-con)}}=\Big{(}\frac{3nH_{0}^{2}\Omega_{0}^{{\rm(c)}}}{2}\Big{)}^{(2+3n)/2},
\end{align}
where the superscript \lq\lq{}{\rm n-con}\rq\rq{} denotes the non-conserved case. Note that this solution is only valid
for $n>0$. Introducing the dynamical
system variables $x=a$ and $y=\dot{a}$, the Raychaudhuri equation (\ref{radius-21}) can be recast as
\begin{align}\label{auto-2}
\dot{y}=-\frac{y^{2}+1}{2x}-\frac{H_{0}^{2}}{2}\Big{(}\Omega_{0}^{{\rm(r)}}x^{-3}-
\frac{3n}{2}\Omega_{0}^{{\rm(c)}}x^{-3\gamma+1}\Big{)}.
\end{align}
This equation together with equation $\dot{x}=y$ construct
a dynamical system with the following eigenvalues
\begin{align}\label{eigen-2}
\lambda^{{\rm(n-con)}}_{1,2}=\pm i\frac{\Big{(}3nH_{0}^{2}\Omega_{0}^{{\rm(c)}}/2\Big{)}^{-(2+3n)/2}}{\sqrt{2+3n}}.
\end{align}
This solution shows that the fixed point $(x=a_{{\rm ES}}^{{\rm(n-con)}}, y=0)$ is a center equilibrium
point. The trajectories of the system are closed curves or cycles winding around the fixed point (see  Fig~\ref{Fig2}).
The evolution of the scale factor versus time has been also depicted
in the left panel of Fig~\ref{Fig3}. The scale factor shows an oscillatory behavior which directly pictures the imaginary nature
of the eigenvalues of the dynamical system (\ref{auto-2}). The right
panel indicates the trajectory in the $(\dot{a},a)$ plane.
The plots of Fig~\ref{Fig3} are drawn for the initial values
$a_{i}=1, \dot{a}_i=0$ and $n=0.66$. This study shows that
when, the condition on conservation of EMT is relaxed,
$f(R,T)$ gravity can contain a stable ES solution.
~
~
\subsubsection{Case iii: $f(R,T)=R+\alpha\chi^{2} T^{\beta}$, $w\neq0,1/3$}\label{sub3}
\begin{figure}[h]
\centering\epsfig{figure=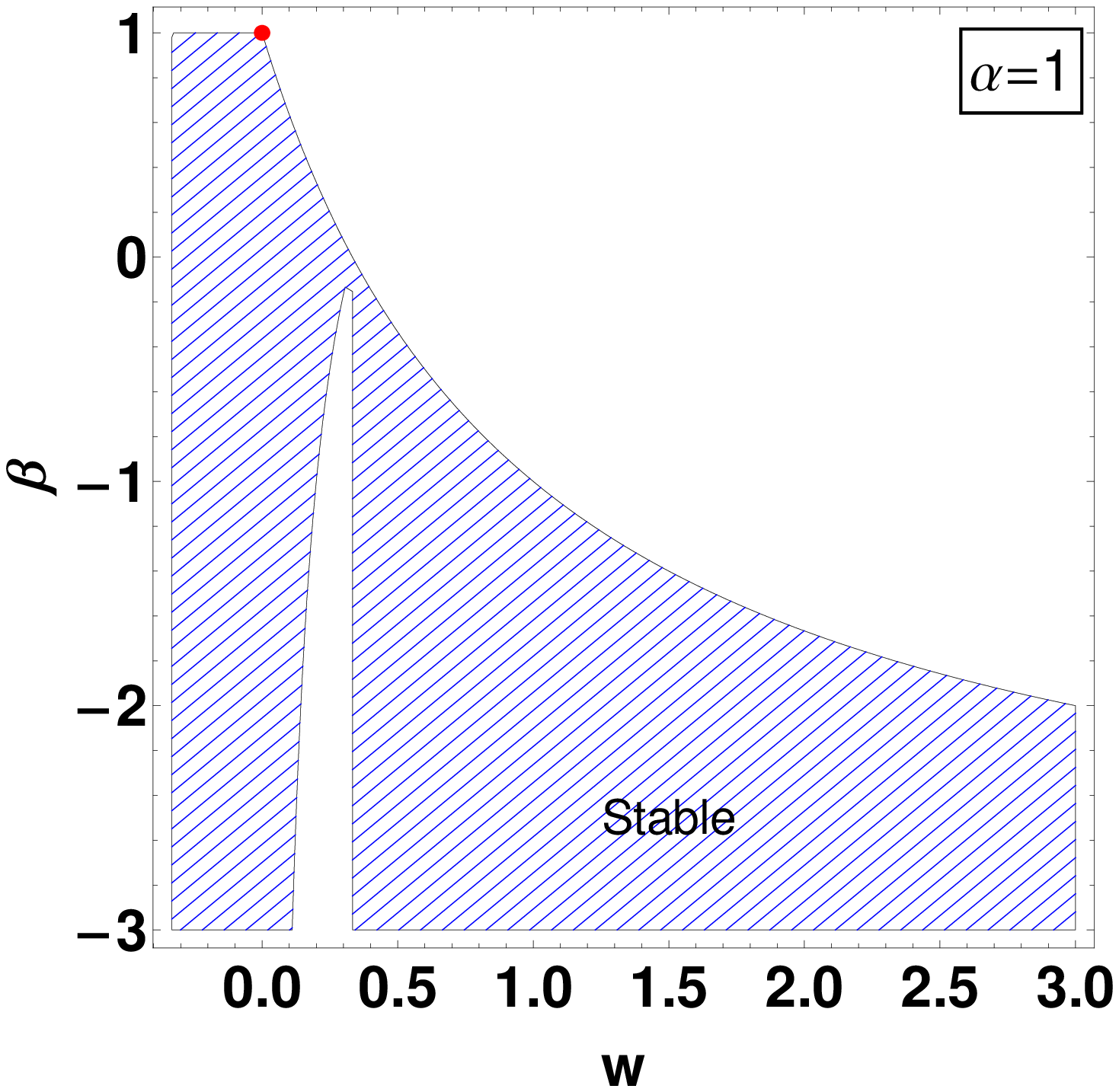,width=5.98cm}
\centering\epsfig{figure=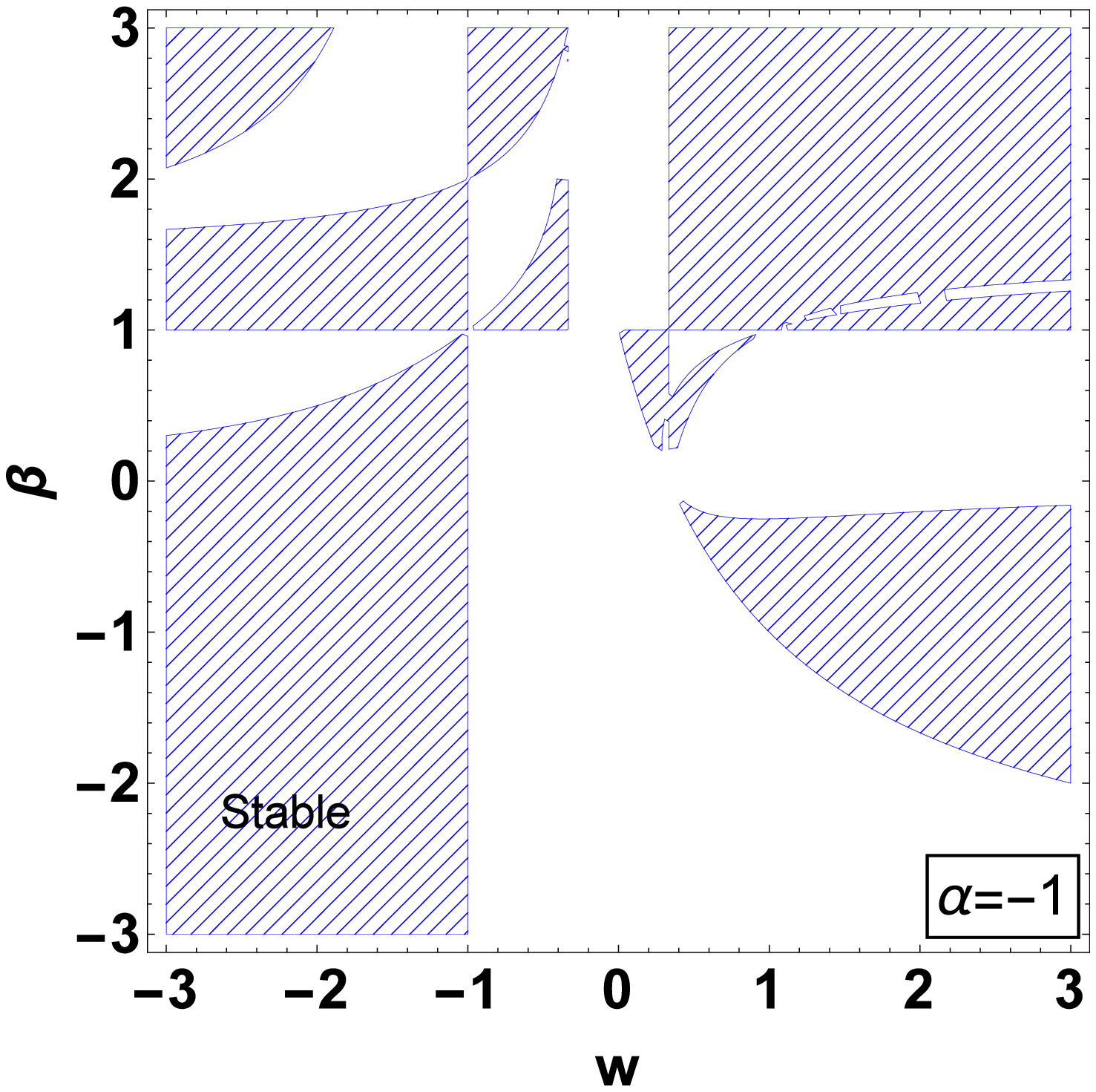,width=5.9cm}
\caption{\footnotesize {The allowed regions for EoS and $\beta$ parameters for which the ES solution is stable. The left figure
is drawn for $\alpha=1$ and the right one is plotted for $\alpha=-1$. In the left figure, red dot denotes the situation of the model $f(R,T)=R+n\chi^{2}T$ with $n=1$.  Note that, for this specific case, the range $n<0$ is not allowed.}}
\label{Fig4}
\end{figure}

For a single barotropic perfect fluid with $p=w\rho$ and choosing
$f(R,T)=R+\alpha\chi^{2} h(T)$, the constraint equation (\ref{relation}) reduces to
\begin{align}\label{relation-w}
\Big{(}1 + \frac{\alpha}{2}(3-w)h' +\alpha (1+w) Th''\Big{)}\dot{T}+
3H(1+w)\Big{(}1 + \alpha h'\Big{)}T=0.
\end{align}
In this case the Friedmann equations can be obtained as follows
\begin{align}\label{Field3-1}
3H^{2}=\chi^{2}\Bigg{\{}\Big{[}1+(1+w)\alpha h'\Big{]}\frac{T}{3w-1}-\frac{\alpha h}{2}\Bigg{\}}-\frac{3}{a^{2}},
\end{align}
and
\begin{align}\label{Field3-2}
2\dot{H}=-\chi^{2}\frac{w+1}{3w-1}(1+\alpha h')T+\frac{2}{a^{2}}.
\end{align}
Eliminating the spatial curvature term, we arrive at an equation in terms
of the Hubble parameter and the trace terms
\begin{align}\label{H-T}
6\dot{H}=-\chi^{2}\Big{(}\frac{3w+1}{3w-1}T+\alpha \frac{w+1}{3w-1}Th'+\alpha h\Big{)}-6H^{2}.
\end{align}
Note that in equations (\ref{relation-w})-(\ref{H-T}), the prime denotes derivative with respect
to the trace and the argument of the function $h(T)$ has been dropped. By determining the function $h(T)$,
equations (\ref{relation-w}) and (\ref{H-T}) make an autonomous system of differential equations for which applying the conditions $\dot{\rho}=\dot{a}=\dot{H}=0$, gives the related critical points. Taking the power law form $h(T)=T^{\beta}$, yields the following values for the scale factor and EMT trace at equilibrium point
\begin{align}
&T_{{\rm ES}}^{{\rm (bar)}}=-\mathfrak{T}^{1/(1-\beta)},\label{criticals}\\
&a_{{\rm ES}}^{{\rm(bar)}}=\Bigg{\{}\frac{6(3w-1)}
{\chi^{2}\mathfrak{T}^{\frac{1}{1-\beta}}\Big{(}2+\alpha\mathfrak{T}^{\beta}
\big{(}1 - 3 w + 2 (1 + w) \beta\big{)}\Big{)}}\Bigg{\}}^{1/2},\label{criticals1}\\
&\mathfrak{T}=\frac{\alpha\Big{(}1-\beta-w(3+\beta)\Big{)}}{1+3w},
\end{align}
where \lq\lq{}{\rm bar}\rq\rq{} denotes the barotropic perfect fluid. The eigenvalues of the system (\ref{relation-w}) and (\ref{H-T}) are obtained as
\begin{align}\label{eig-w}
&\lambda_{1,2}^{{\rm(bar)}}=\pm \frac{X}{Y},\nn
&X=\Bigg{\{}(1+w)(1-9w^{2})(\beta-1)\times\nonumber\\
&\Big{[}2(1-3w)-(1+2w+9w^{2})\beta+2(1+w)(1+3w)\beta^{2}\Big{]}\Big{(}1+w(2\beta-3)\Big{)}\mathfrak{T}^{1/(1-\beta)}\Bigg{\}}^{1/2},\nn
&Y=(3w-1)\Big{[}2(1-3w)-\Big{(}1+2w+9w^{2}\Big{)}\beta+2 (1 + w) (1 + 3 w)\beta^{2}\Big{]},
\end{align}
where we have set $\chi^{2}=1$. The eigenvalues (\ref{eig-w}) imply that the critical point corresponding to (\ref{criticals}) and (\ref{criticals1}) is a saddle one which is an unstable equilibrium point. However, if the expression under curly brackets is set to be negative, the numerator $X$ would be pure imaginary and as a result we have a center equilibrium point. In Fig~\ref{Fig4}, we have sketched the region in the parameter space $(\beta,w)$ that satisfies the condition $X^{2}<0$ (see the shaded zone), for two values $\alpha=\pm1$. Hence, for specific values of $\beta$ and EoS parameters, a stable ES solution could exist.
\par

\begin{figure}[h]
\centering
\includegraphics[scale=0.4]{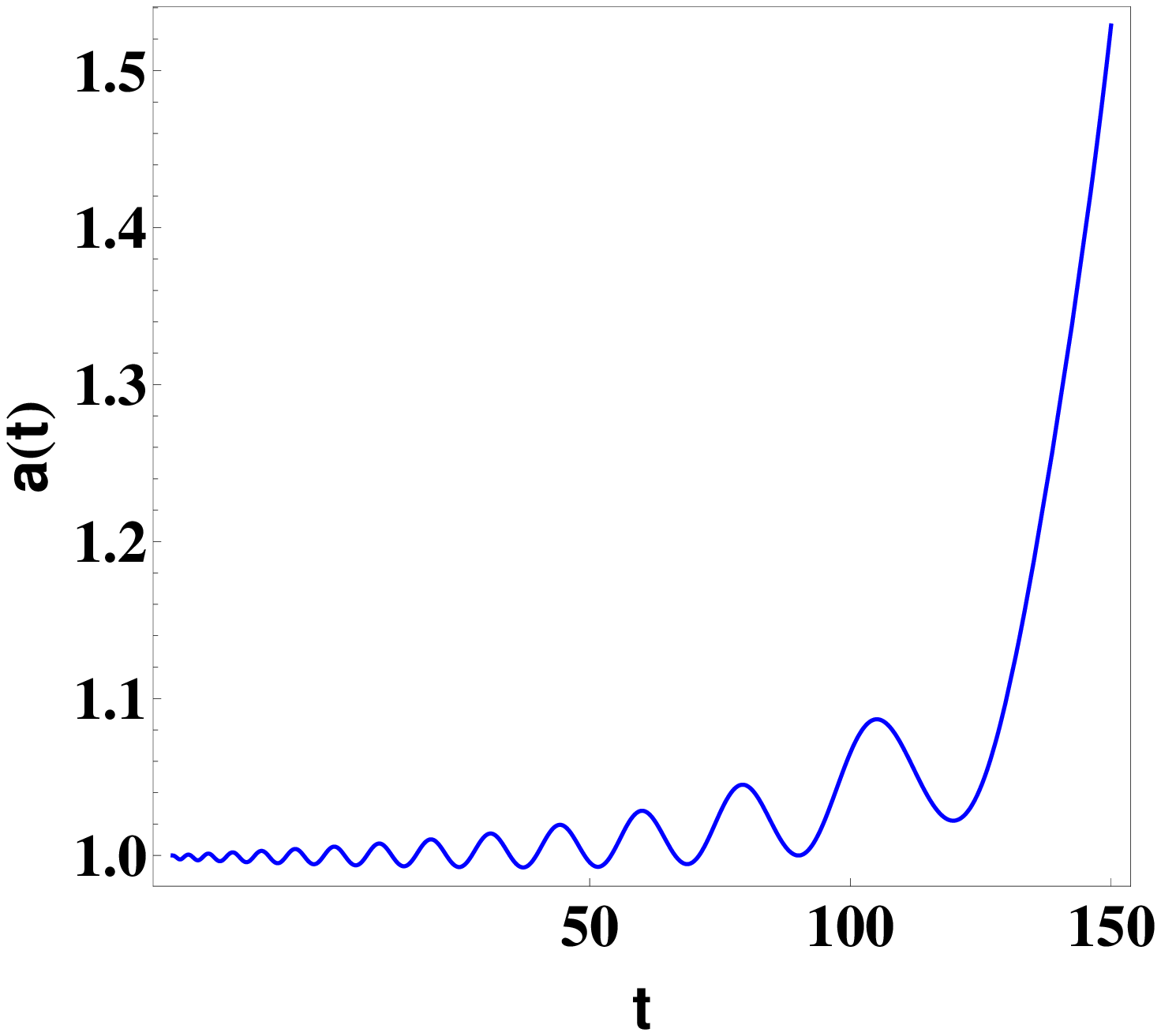}
\includegraphics[scale=0.382]{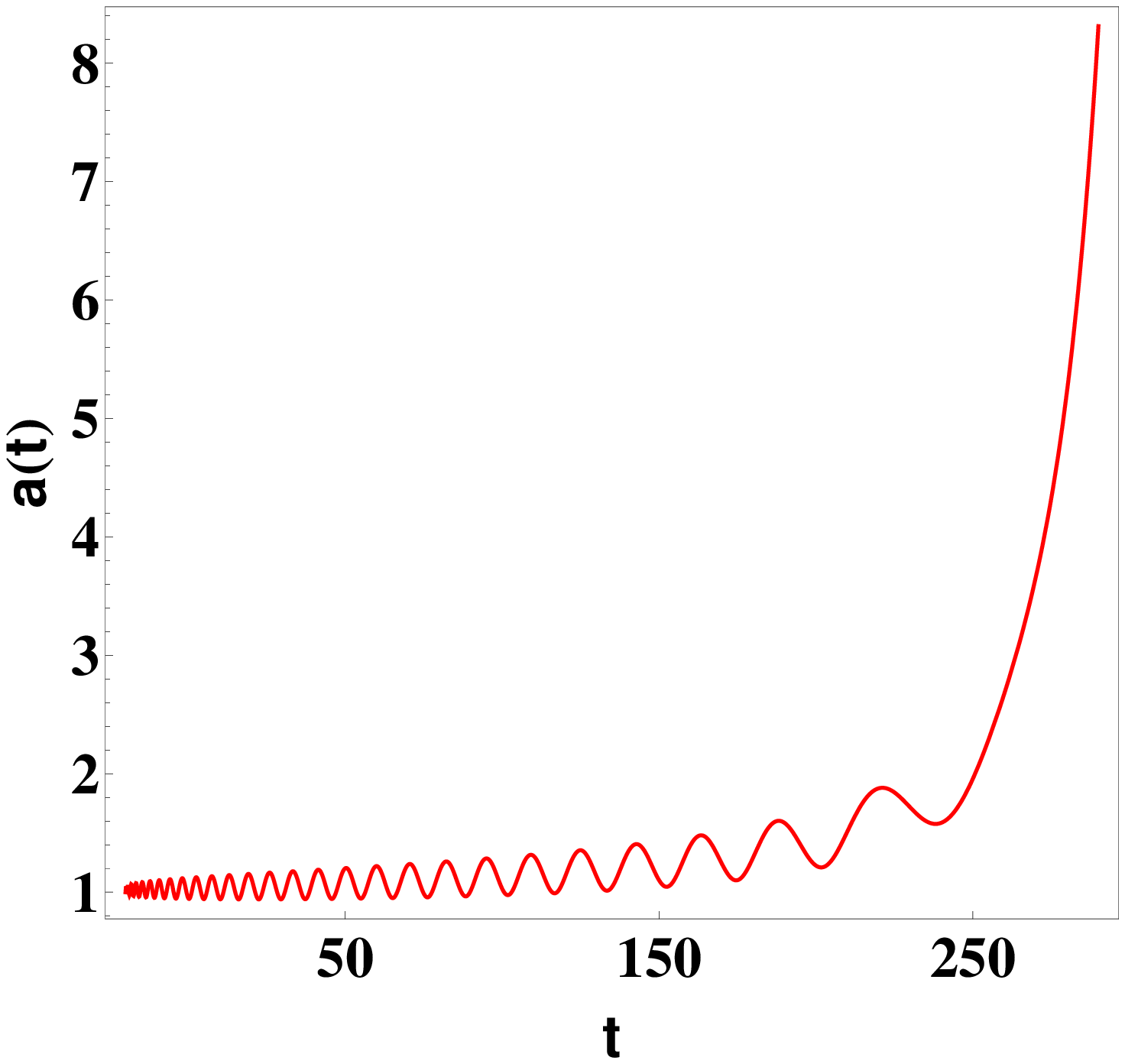}
\caption{\footnotesize {The time behavior of the scale factor in $f(R,T)=R+\chi^{2}\alpha T^{\beta}$, with $\alpha=-1$ and $\beta=5$. In the left panel we have assumed a slowly varying EoS parameter, $w(t)=-1.365+t/410$ with initial conditions $H_{i}=0$ and $T_{i}=-0.828$. The right panel shows the time behavior of the scale factor for $w(t)=0.3-t/601$ with initial conditions  $H_i=0$ and $T{i}=-0.743$.}}
\label{Fig5}
\end{figure}

It is now interesting to examine whether the ES universe presented above could give rise to an EU scenario. To achieve this, we need to check the solutions we have found for case {\rm iii} with more scrutiny.
Let us be more precise. In dealing with equation (\ref{H-T}), depending on the value of the $\beta$ parameter, different roots  can be specified that the physical validity of each root (which corresponds to a scale factor of ES universe) must be verified. Therefore, there can generally exist different fixed points, which correspond to physically acceptable values for $T_{{\rm ES}}$ and $a_{{\rm ES}}$. We note that the corresponding values of $T_{{\rm ES}}$ for the fixed points must be in such a way that the positivity of energy density is preserved, i.e., $\rho_{\rm ES}>0$. Let us now check whether the ES universe presented by case {\rm iii} is capable of providing an EU scenario. Setting $\alpha=-1$ and $\beta=5$ we observe that equation (\ref{H-T}) admits five roots; the first one corresponds to a vanishing value (as a trivial solution) for the trace of EMT which is not physical. The second and third roots are given by, $T_{{\rm ES}}=\pm [(1 + 3 w)/4(1+2w)]^{1/4}$ and the next two are complex conjugates of these roots. Substituting for the negative root into equation (\ref{Field3-1}) leads to the following value for the scale factor at equilibrium point
\begin{align}\label{conn-a}
&a_{{\rm ES}}=\frac{2^{7/4}}{\chi}\left[\frac{(1 + 2 w)^{1/4} (-1 + w + 6 w^2)}{(1 + 3 w)^{1/4} (1 + 8 w + 7 w^2)}\right]^{1/2},\nonumber\\
&T_{{\rm ES}}=-\left[\frac{1 + 3 w}{4(1+2w)}\right]^{1/4},~~~~w < -1,~~ -\frac{1}{3} < w < -\frac{1}{7},
\end{align}
for which, the physical conditions $\rho_{{\rm ES}}>0$ and $a_{{\rm ES}}>0$ will be satisfied within the specified interval for the EoS parameter. Calculations show that the eigenvalues for this fixed point read
\begin{align}\label{conn-eigen}
\lambda_{1,2}=\pm \left[\frac{2 \sqrt{2} (1 + w) (1 + 3 w)^{5/4} (1 + 7 w)}{(1 + 2 w)^{1/4} (-1 + 3 w) (47 + w (184 + 105 w))}\right]^{1/2},
\end{align}
with the following stability properties
\begin{align}\label{conn-stable}
& -\frac{1}{105} (92 + \sqrt{3529}) < w < -1,~~\,{\rm or}\nonumber\\
& -\frac{1}{3} < w <  \frac{1}{105} (-92 + \sqrt{3529}),~~ {\rm or}~~- \frac{1}{7} < w <  \frac{1}{3},~~~ \mbox{a center equilibrium point,}
\end{align}
and
\begin{align}\label{conn-unstable}
& w<-\frac{1}{105} (92 + \sqrt{3529}),~~\,{\rm or}~~~~ -1< w < -\frac{1}{2} ,\nn&~~{\rm or}~~ - \frac{1}{105} (-92 + \sqrt{3529}) < w < -\frac{1}{7},~~{\rm or}~~ w>\frac{1}{3},~~~ \mbox{an unstable point.}
\end{align}
Comparing the ranges of physical validity of $w$, as given by solution (\ref{conn-a}), with the ones specified in (\ref{conn-stable}) and (\ref{conn-unstable}), indicates that the nature of the fixed point depends crucially on the value that the EoS parameter assumes. Such a behavior could be helpful for implementing a cosmological scenario in which, assuming a slowly varying EoS parameter for a short time interval, the universe that has been living in a stable past-eternal static state (a center equilibrium point) could eventually enter into a phase where the stability of the solution is broken leading to an inflationary era (an unstable point). To better illustrate the situation, we investigate two possible cases for a time varying EoS parameter. Let us take $w(t)=-1.365+t/410$ for which the evolution of the scale factor of the universe is plotted in the left panel of Fig.~\ref{Fig5}. It is seen that the universe has started its evolution from an ES state with matter content that the EoS of which is that of a phantom-like matter. As time passes, the purely imaginary eigenvalue of the dynamical system changes to a real value where the center equilibrium point turns into an unstable point. As a result, the universe goes out of the oscillatory phase and enters an inflationary regime. The right panel shows the dynamics of the scale factor for a slowly varying EoS parameter given as, $w(t)=0.28-t/500$. We observe that, having experienced an ES phase, the universe with a radiation-like matter content eventually evolves from such a static phase to an inflationary stage therefore providing an EU scenario for the present $f(R,T)$ model. 

\section{Evolution of the scale factor perturbations in ES universe in the $f(R,T)$ gravity background}\label{pert}
In the present section, we consider linear homogeneous scalar perturbations about the
ES universe and investigate the conditions on its stability against such perturbations. We will therefore extract the governing equation for the evolution of the scale factor perturbations up to the first order. In order to show the consistency of the achieved results, we compare them to the solutions presented in subsections \ref{sub1} and \ref{sub2}. Let us begin with the expressions for perturbed scale factor and energy density, which owing to the homogeneity depend only on time\footnote{Note that all unperturbed quantities are estimated about the equilibrium state at which $a=a_{{\rm ES}}$ and $\rho=\rho_{{\rm ES}}$ and we have dropped the subscript {\rm ES} for convenience.}
\begin{align}\label{pert1}
a(t)=a(1+\delta a(t)),~~~~~~~~~~~~~~\rho(t)=\rho(1+\delta \rho(t)).
\end{align}
Introducing the above relations into the field equation (\ref{fRT field equations}), we seek for the equation that governs the evolution of scale factor perturbation. We further note that the perturbed field equation is linearized and the unperturbed terms will be eliminated to finally have only first order terms. This shall be done using the following background equations of motion
\begin{align}
&f(R,T)=2\frac{\chi^{2}+(1+w)\mathcal{F}}{3w-1}T,\label{unpert1}\\
&F(R,T)=a^{2}\frac{(1+w)(\chi^{2}+\mathcal{F})}{2(3w-1)}T,\label{unpert2}
\end{align}
so that in each equation, substituting $\mathcal{F}=0$ yields the corresponding $f(R)$ gravity model and to obtain GR equations, one must set $F=1$. Using expressions (\ref{pert1}), we obtain perturbations for the Ricci scalar and trace of EMT as
\begin{align}
&\delta R=6\left(\delta a''-2\frac{\delta a}{a^{2}}\right),\label{pert2}\\
&\delta T=(3w-1)\rho \delta \rho=T\delta \rho.\label{pert3}
\end{align}
Substituting equations (\ref{pert1}), (\ref{pert2}) and (\ref{pert3}) into field equations (\ref{fRT field equations}) leads to the perturbed field equations in FLRW background that the tt-component of which reads
\begin{align}\label{pert4}
&\delta \rho=\mathfrak{A}\delta a+\mathfrak{B}\delta a'',\\
&\mathfrak{A}=\frac{6(1+w)\left[-(\chi^{2}+\mathcal{F})+4a^{-2}\mathcal{F}_{,R}\right]}{2\chi^{2}+(3-w)\mathcal{F}+T\mathcal{F}_{,T}},\nonumber\\
&\mathfrak{B}=\frac{-12(1+w)\mathcal{F}_{,R}}{2\chi^{2}+(3-w)\mathcal{F}+T\mathcal{F}_{,T}}.\nonumber
\end{align}
The above equation can be used to eliminate $\delta \rho$ terms that appear in the spatial component of the perturbed field equations. A straightforward but lengthy calculation gives the following evolutionary equation for the scale factor perturbation
\begin{align}\label{pert5}
&\mathcal{A}\delta a(t)+\mathcal{B}\delta a''(t)+\mathcal{C} \delta a^{(4)}(t)=0,\\
&\mathcal{A}=\frac{1+w}{3w-1}(\chi^{2}+\mathcal{F})T-\frac{24}{a^{4}}F_{,R}+\mathfrak{A}\left[\frac{2}{a^{2}}TF_{,T}-\frac{1}{2}T\mathcal{F}-\chi^{2}\frac{w}{3w-1}T\right],\nonumber\\
&\mathcal{B}=-2F+\frac{24}{a^{4}}F_{,R}-\mathfrak{A}TF_{,T}+\mathfrak{B}\left[\frac{2}{a^{2}}TF_{,T}-\frac{1}{2}T\mathcal{F}-\chi^{2}\frac{w}{3w-1}T\right],\nonumber\\
&\mathcal{C}=-6F_{,R}-\mathfrak{B}TF_{,T},~~~~X(R,T)_{,T}\equiv\partial X/\partial T,~~~~~Y(R,T)_{,R}\equiv\partial Y/\partial R.\nonumber
\end{align}
Equation (\ref{pert5}) is the most general equation for the scale factor perturbation around the ES state and to exploit its predictions an underlying model must be determined.\par

\begin{figure}[tbp]
\centering
\includegraphics[scale=0.395]{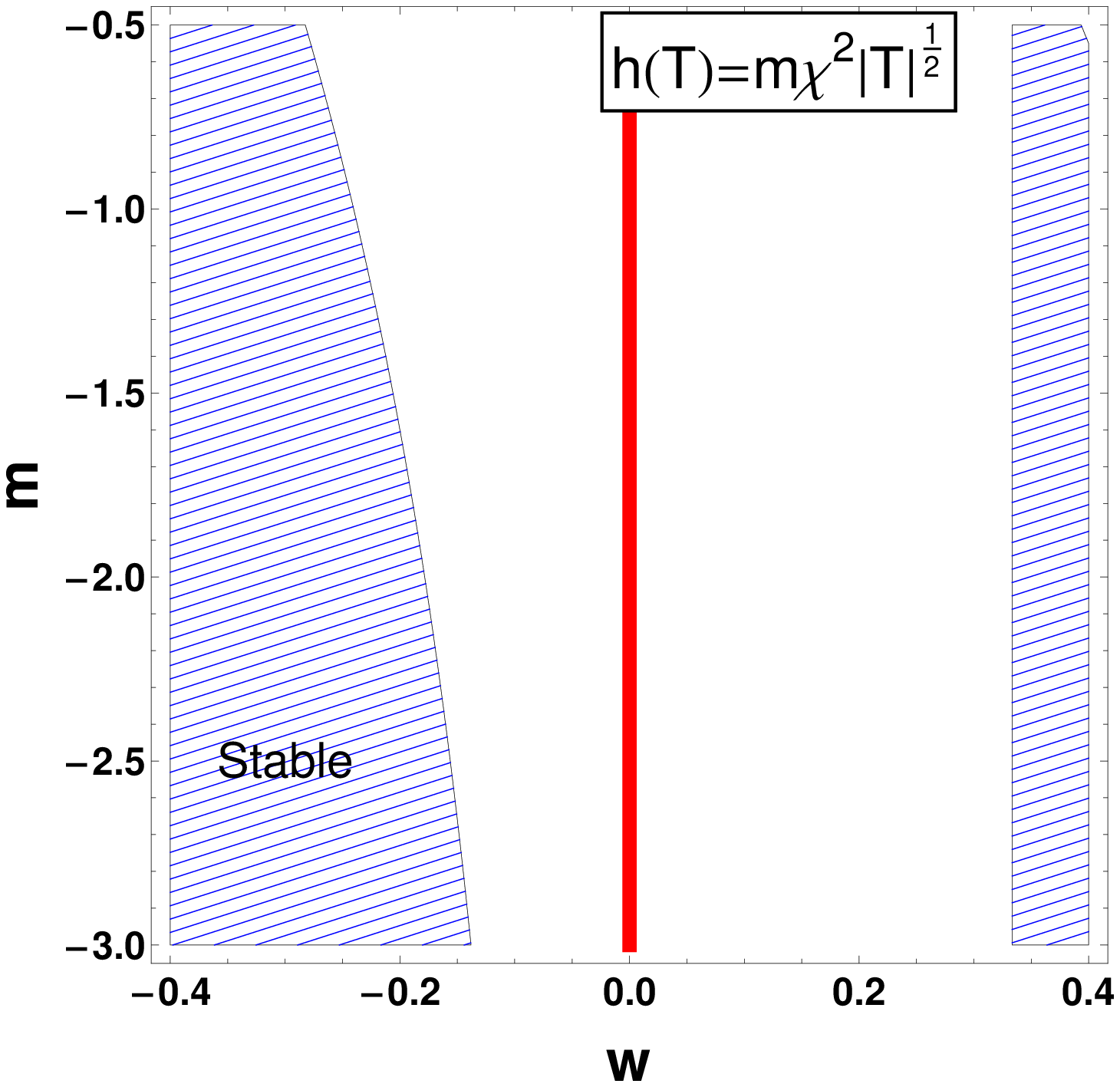}
\includegraphics[scale=0.385]{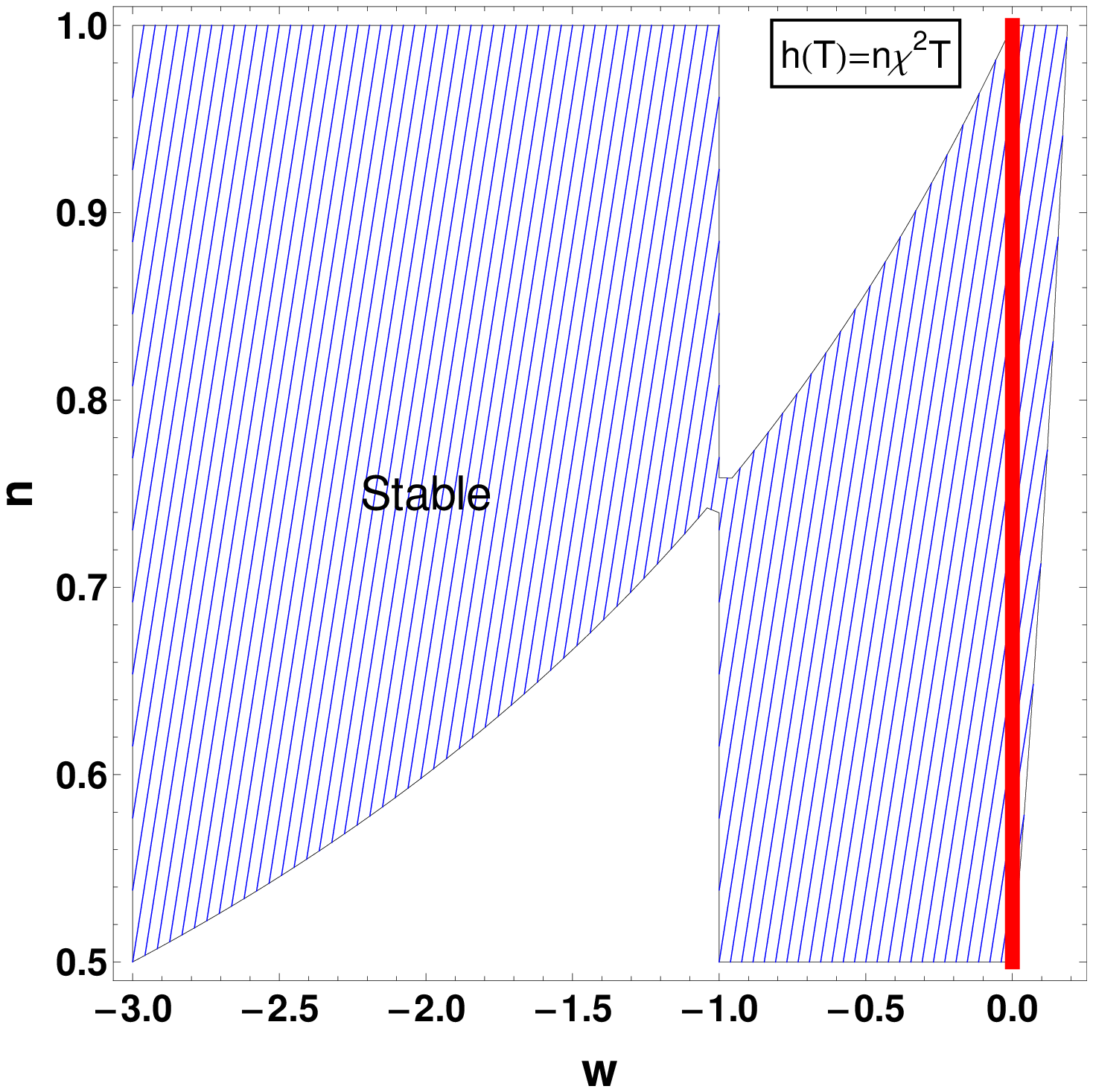}
\caption{\footnotesize {The stable regions in the 2D spaces constructed out of the coupling constants $m$, $n$ and the EoS parameter $w$, for the models with $h(T)=m\chi^{2}\sqrt{T}$ and $h(T)=n\chi^{2}T$}, respectively. Red lines show the solutions corresponding to $w=0$ in each model.}
\label{Fig6}
\end{figure}

Next, we discuss differential equation (\ref{pert5}) for GR and two cases which are considered in subsections \ref{sub1} and \ref{sub2}.  For GR we have $F=1$ and $\mathcal{F}=0$, so that equation (\ref{pert5}) reduces to the following equation
\begin{align}\label{GR}
2\delta a''-\rho(1+w)(1+3w)\delta a=0.
\end{align}
This result and the corresponding solution have been also reported in~\cite{fR pert1}. Because of the forth order nature of equation (\ref{pert5}), we take the following ansatz as the solution
\begin{align}\label{ansatz}
\delta a(t)=C_{1}e^{\omega_{1}t}+C_{2}e^{-\omega_{1}t}+C_{3}e^{\omega_{2}t}+C_{4}e^{-\omega_{2}t},
\end{align}
whereby the frequencies $\omega_{1}$ and $\omega_{2}$ read
\begin{align}\label{sol-ansatz}
\omega_{i}=\left[\frac{-\mathcal{B}\pm\sqrt{\mathcal{B}^{2}-4\mathcal{A}\mathcal{C}}}{2\mathcal{C}}\right]^{\f{1}{2}},~~~~~ i=1,2.
\end{align}
We therefore observe that depending on different types of $f(R,T)$ models and the free parameters exploited, the ES universe could be stable or unstable against homogeneous perturbations. The models that we have considered in subsections \ref{sub1} and \ref{sub2} are in the form $f(R,T)=R+h(T)$ for which we have $\mathcal{B}=-2$ and $\mathcal{C}=0$. Thus, for these two models, the differential equation governing the scale factor perturbations is similar to the GR case but with different frequency, given as
\be\label{omega12}
\omega=\omega_1=\omega_2=\sqrt{\mathcal{A}/2},~~~~C_2=C_3=0.
\ee
Therefore, the criterion for a stable solution would be $\mathcal{A}<0$, otherwise, the perturbation in scale factor will diverge. Some algebraic calculations for the model $f(R,T)=R+m\chi^{2}\sqrt{T}$ reveal that
\begin{align}\label{pert-modelI}
\mathcal{A}=\frac{2 (1 + w) (-2 + m + 6 w) (-1 + 2 m w + 9 w^2)}{(1 - 3 w) (4 - 3 m - 12 w + m w)}\chi^{2}\rho.
\end{align}
The above expression for conserved case with $w=0$ reduces to $2 ( m-2)/(3m-4)$ from which we see that the conserved case amounts to an unstable ES solution since $\mathcal{A}>0$ for all valid values of the parameter $m$, i.e., for $m<0$. For non-conserved model with $f(R,T)=R+n\chi^{2}T$, we get
\begin{align}\label{pert-modelII}
\mathcal{A}=(w+1) (n + 1)\left[1 + \frac{3 n (3 w - 1) + 6 w}{3 - n (3 - w)}\right]\chi^{2}\rho,
\end{align}
which for the special case $w=0$ reduces to $(1 - 2 n) (1 + n)/(1 -  n)$. We then conclude that, the condition for stability i.e., $ \mathcal{A}<0$ is satisfied for $1/2 < n < 1$. This interval for $n$ parameter respects the condition on physical validity of the radius of ES universe, as required by equation (\ref{sol-radius-2}). In Fig~\ref{Fig6}, stable regions for both models have been drawn. The solution corresponding to $w=0$ is indicated by a red line.

\section{Concluding remarks}\label{con}
In this work we have studied the existence and stability of
the ES universe in $f(R,T)$ modified gravity theories. In these theories, the Lagrangian of $f(R)$ gravity is extended to include the trace of EMT which in turn would allow for remarkable outcomes in the gravitational interactions. Having employed a homogeneous and isotropic FRLW metric with spatially positive curvature\footnote{This non-zero curvature will become insignificant in the late times and the positivity authorizes the universe to enter an inflationary stage in the early times through which the hot big-bang epoch initiates through a reheating stage \cite{Ellisemergent}.}, we examined the stability of the ES solution with the help of dynamical system approach. The $f(R,T)$ function has been chosen as a linear combination of the Ricci curvature scalar and an arbitrary function of EMT trace, denoted as $h(T)$. Two main classes of models we have established here, include: $f(R,T)$ gravity models which respect the conservation of EMT and those that do not. In the former class, the conservation of EMT results in $h(T)=C_{1}\sqrt{|T|} +C_2$, for a pressure-less fluid. For this case, considering as the matter contents,  a mixture of radiation and cold dark matter, an ES solution can be found. However, the study of the trajectories of the related dynamical system near the critical point suggests that the ES solution is unstable (of the saddle type) for all valid values of the coefficient $C_{1}$. Therefore, this class of models cannot be served as a stable ES solution, so that like in the GR case \cite{Ellisemergent}, a fine-tuning is required if the ES state is to be the initial state of the universe for a past-eternal inflationary cosmology. In the latter class, by relaxing the condition on EMT conservation and applying Bianchi identity, we have obtained a covariant relation between derivatives of $h(T)$, EMT and its trace. This case has been investigated under two subclasses. In the first one, choosing $h(T)=n\chi^2 T$, we obtained an ES solution which is stable for $n>0$ in the sense that its dynamical behavior corresponds to a center equilibrium point. In the second one, we have considered a perfect fluid with linear EoS ($w=p/\rho$) as the matter content. Choosing the function $h(T)=\alpha\chi^2 T^{\beta}$, we showed that a stable ES universe could indeed exist, in the same sense as the previous case, depending on the values of the parameters $w$, $\alpha$ and $\beta$. Therefore, the second class of solutions suggest that the universe in $f(R,T)$ modified gravity can remain at a stable state past-eternally, and may go through a series of infinite non-singular oscillations around this state. We then conclude that, in contrast to $f(R)$ gravity in which, unstable ES solutions do generally exist~\cite{genericfRstability}, $f(R,T)$ modified gravity could potentially admit stable ES solutions for some specific forms of $f(R,T)$ function.
\par
Finally, as we near to close this paper, there remain two points that beg some additional elucidation. Firstly, the stable ES solutions we have found raise this question that, in order to have a successful cosmological scenario, the regime of infinite cycles around the center equilibrium points must be able to eventually break and then enters the current expanding phase of the universe \cite{branesta1,lqgsta}. This purpose can be achieved by varying one of the model parameters, namely the EoS parameter $w$ and the dimensionless parameter $n$, so that the system could undergo a bifurcation which results in changing the topological structure of the phase space (see e.g., \cite{modifiedgsta1}). In the current study, we observed that the equilibrium point for case {\rm iii} could be stable or unstable depending on the value of $w$ parameter. We then assumed that this parameter changes for a short period of time during which, the center equilibrium point converts to an unstable point or correspondingly the phase of the universe changes from an infinite number of oscillations about the ES state to an inflationary regime. Therefore, $f(R,T)$ models presented here can provide a setting in which an ES universe is connected to an asymptotic EU scenario.
Secondly, in order to be sure that the universe can stay at the static state past-eternally,  therefore allowing for a successful implementation of the emergent scenario, the ES solution must be stable against all types of perturbations. In this regard, we have performed homogeneous and linear scalar perturbations in the scale factor and energy density and it is found that the ES universe is stable against these type of perturbations under a variety of the obtained conditions. It is also of interest to extend our results  to include the inhomogeneous perturbations around the ES state which indeed could provide a richer structure for stability/instability analysis of Einstein cosmos in $f(R,T)$ modified gravity theory. Work along these lines is currently underway.

\end{document}